\documentclass{jfm}
\usepackage{graphicx}
\usepackage{epstopdf, epsfig}
\usepackage{xcolor}
\usepackage{MnSymbol}
\usepackage{wasysym}

\shorttitle{In line wake interaction}
\shortauthor{N. S. Lagopoulos, G.D. Weymouth and B. Ganapathisubramani}

\title{Deflected Wake Interaction of Tandem Flapping Foils}

\author{ N. S. Lagopoulos\aff{1}
  \corresp{\email{N.Lagopoulos@soton.ac.uk}},
  G. D. Weymouth\aff{2}  \corresp{\email{G.D.Weymouth@soton.ac.uk}},
 \and B. Ganapathisubramani\aff{1}}

\affiliation{\aff{1}Aerodynamics and Flight Mechanics Group, University of Southampton, UK
\aff{2}Southampton Marine and Maritime Institute, University of Southampton and Alan Turing Institute, London, UK}

\begin{document}

\maketitle

\begin{abstract}
Symmetric flapping foils are known to produce deflected jets at high frequency-amplitude combinations even at a zero mean angle of attack. This reduces the frequency range of useful propulsive configurations without side force. In this study, we numerically analyse the interaction of these deflected jets for tandem flapping foils undergoing coupled heave to pitch motion in a two dimensional domain. The impact of the flapping Strouhal number, foil spacing and phasing on wake interaction is investigated. Our primary finding is that the back foil is capable of cancelling the wake deflection and mean side force of the front foil, even when located up to 5 chord lengths downstream. This is achieved by attracting the incoming dipoles and disturbing their cohesion within the limits of the back foil's range of flapping motion. We also show that the impact on cycle averaged thrust varies from high augmentation to drag generation depending on the wake patterns downstream of the back foil. These findings provide new insights towards the design of biomimetic tandem propulsors, as they expand their working envelope and ability to rapidly increase or decrease the forward speed by manipulating the size of the shed vortices.

\end{abstract}


\section{Introduction}  \label{intro}

\begin{figure}
  \centerline{\includegraphics[width =5.5 in]{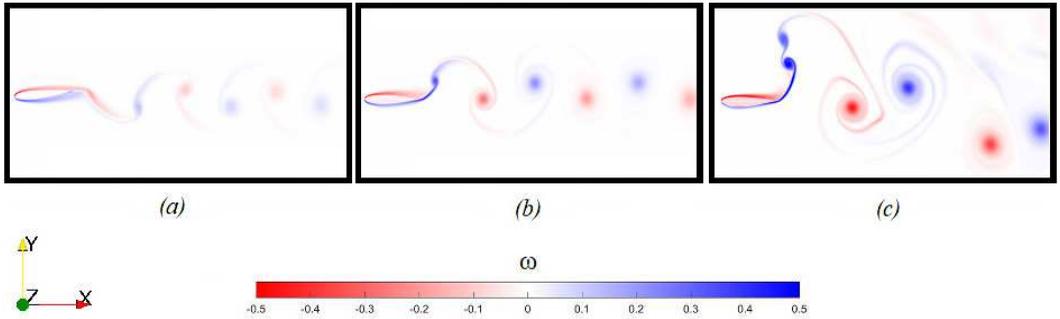}}
  \caption{Wake transitions of a flapping foil undergoing heave to pitch coupling at $St_C=0.625$: (a) $\emph{von}~K\acute{a}rm\acute{a}n$  street at $A_C =0.2$ (b) reversed $\emph{von}~K\acute{a}rm\acute{a}n$  street at $A_C =0.3$ and (c) deflected jet at $A_C =0.4$. }
\label{fig:1}
\end{figure}

Due to their prevalence within the airborne and aquatic wildlife, single flapping foils have caught the interest of scientists and engineers alike since the early twentieth century \citep{Knoller1909, Betz1912}. Moreover, tandem flapping configurations e.g. insect wings  \citep{Alexander1984, Thomas2004}, plesiosaur flippers  \citep{MuscuttEX2017} etc. are shown to outperform single flappers under certain wake-to-wake interactions.

For a single flapping foil the onset of thrust generation is marked by a reverse $\emph{von}~K\acute{a}rm\acute{a}n$  street (see figure \ref{fig:1}b) downstream of its trailing edge (T.E.) \citep{vonKarman1935}, although a lag between the two conditions exist \citep{Diana2008b, Bohl2009, Lagopoulos2019}. This wake pattern is determined by the oscillating T.E. amplitude $A$ and frequency $f$ of the motion, which form together an $amplitude$ $based$ Strouhal number $St_A = (2fA)/U_{\infty}$ as described by \cite{Triantafyllou1991}. An increasing $St_A$ leads to permanent deflection of the jet (see figure \ref{fig:1}c) and thus side force generation even when both the camber and the mean angle of attack are zero \citep{Diana2008a, Cleaver2012}. This is the result of $dipole$ formation when shedding vortices become strong enough to attract each other and depart from the centreline \citep{Diana2008a, Diana2008b}. Although three dimensional effects compromise the coherence of these structures \citep{Zurman2020}, the formation and subsequent deflection of the dipole maintains its quasi two dimensional nature \citep{Couder1986, Diana2008b}.  

To improve the propulsive performance of a flapping system, various researchers have proposed the use of multiple foil configurations. In particular, tandem flapping foils are shown to improve thrust generation via $wake$ $recapture$ both numerically \citep{Muscutt2017,Broering2012,Akhtar2007} and experimentally \citep{MuscuttEX2017,Warkentin2007,Usherwood2008}. More specifically, thrust and efficiency augmentation can be achieved when the hind foil is weaving within the incoming vortices shed by the front one, determined by the inter foil spacing and phase lag.

A prominent feature of these studies is the relative lack of influence the downstream foil is said to have on the wake and forces of the upstream foil. However, those studies focus on cases with symmetric reverse $\emph{von}~K\acute{a}rm\acute{a}n$ streets where momentum exchange between shedding vortices is minimal. On the other hand, vortices of deflected wakes are often in very close proximity to each other, forming a long chain of well defined and correlated dipoles. 

This paper focuses on the interaction between deflected wakes of tandem flapping foils undergoing harmonic motion. Simulations are conducted for single and tandem flapping configurations at a range of $St_{A}$ that ensures steady dipole formation. As the primary mechanism for wake deflection is two dimensional, we restrict ourselves to two-dimensional simulations in this work. It is revealed that certain phase-spacing combinations neutralise deflection for both foils. Distinct types of wake to wake interaction are observed and described in terms of their propulsive characteristics. In addition we clarify the mechanism of deflection cancellation and determine its limits in terms of a simple non-dimensional parameter, the spacing based Strouhal number. 

\label{sec:types_paper}
\section{Methodology}
\subsection{Geometry and kinematics}

\begin{figure}
  \centerline{\includegraphics[width =3.2 in]{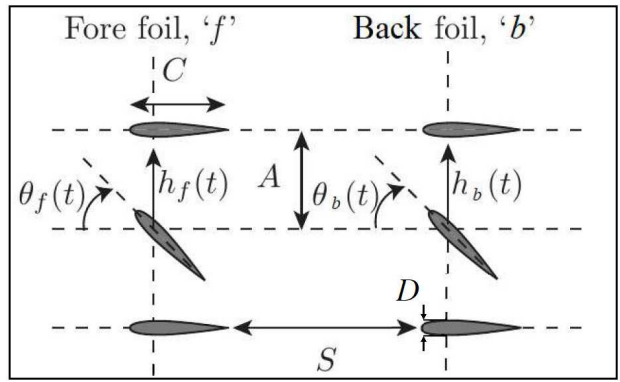}}
  \caption{The kinematic parameters of a tandem configuration. Redrawn from \cite{Muscutt2017}.}
\label{fig:2}
\end{figure}

Figure \ref{fig:2} shows the basic elements of the simulated configuration. Two rigid NACA0016 with a thickness $D = 0.16~ \mathcal{C}$ (where $\mathcal{C}$ is the chord length) undergo sinusoidal heave to pitch coupling around the quarter chord, against a uniform free stream velocity $U_{\infty}$. The pitching component of the motion is a sinusoidal rotation about the pivot point $\mathcal{P} = 0.25$ (normalised by the chord $\mathcal{C}$) while the heaving part is a sinusoidal vertical translation with respect to the centreline. Since both motions are harmonic, coupling is simply achieved by the superposition of the two kinematic components: 

\begin{equation}
\begin{array}{l}
\displaystyle ~~~~~~~~~~~~~~~~~~~~~y_s(t) = y_ h(t) + y_ {\theta}(t)~~~where:\\
\\
\displaystyle  y_h(t) =  h_\mathit{0} \sin(2f\pi t)~~~~,~~~~y_ {\theta}(t) =(1-\mathcal{P})\mathcal{C}\sin({\theta}(t))~~~and:\\
\\
\displaystyle~~~~~~~~~~~~~~~~~~~~~\theta(t) = \theta_\mathit{0} \sin(2f\pi t +\psi)\\ 
\end{array} 
\label{eq:xdef}
\end{equation}
where subscript $s$ refers to the single foil configuration while the subscripts $h$ and $\theta$ refer to the heaving and pitching component respectively. 

The instantaneous angle of the foil due to pitching is expressed as the harmonic $\theta(t)$ with the $h_{0}$ and $\theta_{0}$ being the amplitudes of pure heave and pure pitch respectively. In addition, the heave to pitch phase difference is set as $\psi$ = $90^{\circ}$, which is shown to maximize the propulsive efficiency of the foil within the frequency range of interest \citep{Platzer2008}.

Another important kinematic parameter in coupled motions is the effective angle of attack $\alpha_\emph{eff}(t)$, which is the summation of the instantaneous pitch angle $\theta(t)$ and the heave induced angle of attack. Thus for $\psi$ = $90^{\circ}$ the amplitude of $\alpha_\emph{eff}(t)$ is:

\begin{equation}
\alpha = \arctan \frac{2 \pi f h_{0}}{U_\infty} - \theta_{0}
\end{equation}
where $2 \pi f h_{0}$ is the amplitude of $dy_h/dt$. Here, we set $\alpha=10^{\circ}$ to achieve high efficiency kinematics \citep{Muscutt2017}. Furthermore, the phase lag between the  two foils is expressed as $\varphi$ and will be referred to as simply the $phase$: 

\begin{equation}
\begin{array}{l}
\displaystyle ~~~~~~~~~~~~~~y_f(t) = y_ h(t) + y_ {\theta}(t)\\
\\
\displaystyle ~~~~~~~~~~~~~~y_b(t) = y_ h(t+ \varphi) + y_ {\theta}(t + \varphi)\\
\end{array}
\end{equation}
where subscripts $f$ and $b$ denote the front and back foils respectively.

\subsection{Dimensionless parameters}
Three aerodynamic parameters, scaled by $\mathcal{C}$, are used to describe the interaction between the flapping foil and the free stream \citep{Koochesfahani1989, Marais2012, Kim2019}. These are the Reynolds number $Re=U_{\infty} \mathcal{C} / \nu$ (with $\nu$ being the kinematic viscosity), the Strouhal number, $St_c$ and the normalised peak-to-peak T.E. amplitude, $A_C$ defined as:
\begin{equation}
St_C = \frac{f~\mathcal{C}}{U_{\infty}}~~,~~A_C =\frac{2 A}{\mathcal{C}}
\end{equation}
where $A$ is the cumulative T.E. amplitude of the coupled motion.  Note that the use of $\mathcal{C}$ in the definition of the above parameters enables us to retrieve the classic amplitude based Strouhal via the product $St_C \cdot A_C =St_{A}$. The latter can be understood as the ratio between the speed of the foil tip and $U_{\infty}$ \citep{Diana2008b}. To allow comparison with the results of \cite{Diana2008b}, all simulations of this study are conducted for $Re=1173$ . 

Another key parameter of a tandem configuration is the distance between the T.E. of the front foil and the L.E. of the back foil on the streamwise direction known as the inter foil $spacing$ $\mathcal{S}$ (see figure \ref{fig:2}). Therefore the chord normalised spacing $\mathcal{S}_C$ is defined as:

\begin{equation}
\mathcal{S}_C = \frac{\mathcal{S}}{\mathcal{C}}
\end{equation}

The propulsive performance of the system is characterised by the thrust and lift coefficients. These are the x and y components of the force acting on the foil, normalised by the dynamic pressure:

\begin{equation}
C_t = \frac{F_x}{\frac{1}{2}\rho U_\infty ^2 \mathcal{C}} ~~,~~C_l = \frac{F_y}{\frac{1}{2}\rho U_\infty ^2 \mathcal{C}}
\end{equation}
Cycle averaged quantities are presented with an overbar to distinguish them from their instantaneous counterparts. 
\subsection{Computational method}\label{sec:filetypes}
The CFD solver utilized in this work is capable of simulating complex geometries and moving
boundaries for a variety of Reynolds numbers in 2D and 3D domains, via
the boundary data immersion method BDIM \citep{Schlanderer2017}. BDIM solves
the viscous time-dependent Navier-Stokes equations and simulates the entire domain
by combining the moving body and the ambient fluid through a kernel function. This
technique has quadratic convergence and has been validated for flapping foil simulations over a wide span of kinematics \citep{Maertens2015,Polet2015}.

The mesh profile is a rectangular Cartesian grid. A dense uniform grid is used near the body and in the near wake while an exponentially stretched mesh is used in the far-field. The boundary conditions consist of a uniform inflow, zero-gradient outflow and free-slip conditions on the upper and lower boundaries. Moreover, no slip conditions are imposed on the surface of the oscillating foil. 

Mesh density is indicated by the number of grid points per chord. A grid convergence analysis was conducted to identify the appropriate resolution. As documented in \cite{Lagopoulos2019}, a grid spacing of  $\delta x = \delta y = \mathcal{C}/192$ results in force predictions with less than $3\%$ error compared to a grid with twice the resolution in each direction and is therefore used for all simulations in this manuscript. 

\section{Results and discussion}
\subsection{Single foil analysis}

Single foil arrangements undergoing harmonic heave to pitch coupling are tested for $St_C$ $\sim$ $[0.625~,~2.5]$ and $Re=1173$ in a two dimensional domain. $A_C$ is chosen so that $\overline{C}_{l,s} \sim 0.4=const.$ across the entire $St_C$ range, ensuring steady wake deflection. Furthermore, the relatively high Strouhals guarantee that, all deflected wakes are thrust producing. The kinematic details and resulting $\overline{C}_{l,s}$ are shown in table \ref{tab:1}.

\begin{figure}
  \centerline{\includegraphics[width =5.6 in]{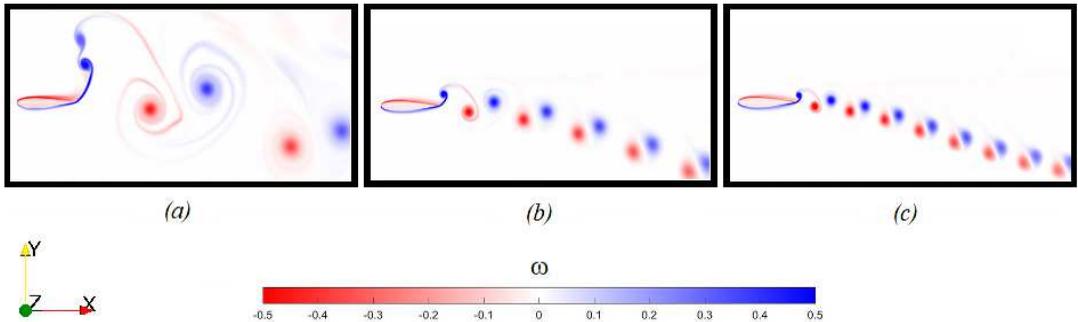}}
  \caption{Normalised vorticity field of a single flapping foil undergoing harmonic heave to pitch coupling at (a) $St_C=0.625$, (b) $St_C=1.5625$ and (c) $St_C=2.5$. All produced wakes are steadily deflected generating $\overline{C}_{l,s} \sim 0.4$.}
\label{fig:3}
\end{figure}

\begin{table}
\begin{center}
\def~{\hphantom{0}}
 \begin{tabular}{c c c c c c c} 
 $St_C$ & $~~$ & $A_C$ & $~~$ & $\overline{C}_{t,s}$ & $~~$ &$\overline{C}_{l,s}$ \\ [3pt] 
  0.6250 &  & 1.405 &  & 0.59 &  & 0.400\\
  0.9375 &  & 0.651 &  & 0.49 &  & 0.400\\ 
  1.2500 &  & 0.437 &  & 0.42 &  & 0.400\\
  1.5625 &  & 0.326 &  & 0.37 &  & 0.400\\
  1.8750 &  & 0.258 &  & 0.32 &  & 0.390\\ 
  2.1875 &  & 0.213 &  & 0.28 &  & 0.395\\ 
  2.5000 &  & 0.181 &  & 0.33 &  & 0.405\\ 
  \end{tabular}
 \caption{Kinematics of a single flapping foil undergoing coupled heave and pitch motions with $\alpha  = 10^\circ$ and  $\psi = 90^\circ$. All cases are deflected and the amplitude $A_C$ has been tuned to achieve nearly identical $\overline{C}_{l,s}$ across the $St_C$ range.}
\label{tab:1}
\end{center}
\end{table}

The resultant wakes (see figure \ref{fig:3}) maintain the basic features and deflection mechanism of asymmetric jets reported in literature  for pure pitch \citep{Diana2008a,Diana2008b, He2012} and pure heave \citep{Cleaver2012, Kozlowski2014}. Vortex circulation $\Gamma$ is proportional to the flapping frequency while the opposite is true for the distance between consecutive vortices. In the beginning, the first shed vortex follows an independent path away from the centreline. Yet, the distance between the second and the third subsequent vortices is noticeably smaller. This results in the formation of a dipole as shorter distances lead to stronger synergy among vortices according to the $Biot-Savart$ vortex induction law \citep{Zheng2012}. The initial dipole departs from the centerline, breaking the symmetry of the mean jet and imposing its path to the subsequent dipole \citep{Diana2008b}. 
\subsection{Tandem foil analysis}
Two dimensional tandem foils are tested for the kinematics and $Re$ of the previous section. Simulations are performed for a wide range of spacings $\mathcal{S}_C$ $\sim$ $[1 -6]$ and $\varphi \sim [0\pi - 1.75 \pi]$ in increments of $1\pi/8$, respectively. It is revealed that, at certain $\mathcal{S}_C$ - $\varphi$ combinations, the presence of the downstream foil results in a stable symmetric wake and zero net lift, even for the upstream foil. Specifically, we define lift cancellation as the condition when:

\begin{equation}
\begin{array}{l}
\displaystyle ~~~ |{\overline{C_{l,f} }}| < \epsilon ~~~~,~~~~ |{\overline{C_{l,b} }}| < \epsilon ~~~~,~~~~|{\overline{C_{l,f}} + \overline{C_{l,b} }}| < \epsilon\\
\\
\displaystyle  ~~~~~~~~~~~~~~~~ where ~~~~~\epsilon = 0.05~\overline{C_{l,s}}\\
\\
\end{array} 
\label{eq:xdef}
\end{equation}

\begin{figure}
  \centerline{\includegraphics[width =4.7 in]{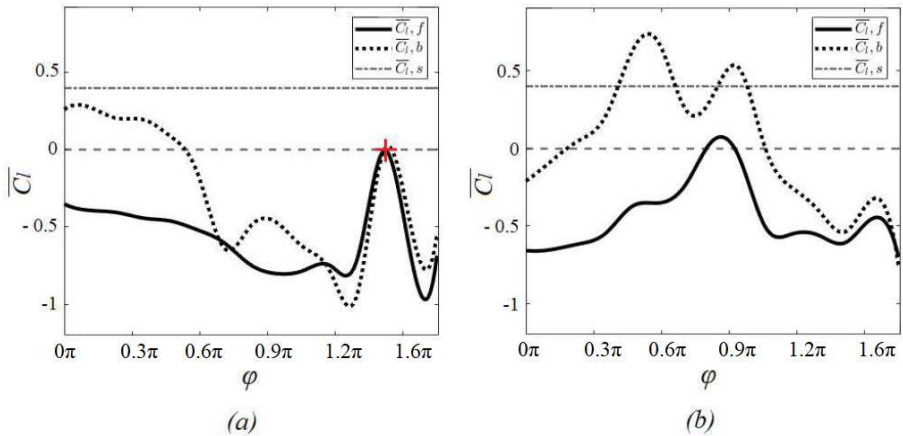}}
  \caption{Cycle averaged Lift coefficient versus phase for a tandem foil configuration at $St_C =1.25$ and (a) $\mathcal{S}_C =3$ , (b) $\mathcal{S}_C =4$. The red cross in plot (a) marks lift cancellation.}
\label{fig:4}
\end{figure}

This condition is illustrated for two spacing at $St_C=1.25$ in figure \ref{fig:4}. At $\mathcal{S}_C=3$ and $\varphi=1.5\pi$ both $\overline{C}_l$ curves are converging to zero, demonstrating complete lift cancellation on the front and back foil due to wake interaction. However, at $\mathcal{S}_C=4$, while at least two phases lead to $\overline{C}_{l,f} =0$, there is no phase which causes lift cancellation on both foils.  

\begin{figure}
  \centerline{\includegraphics[width =6.2 in]{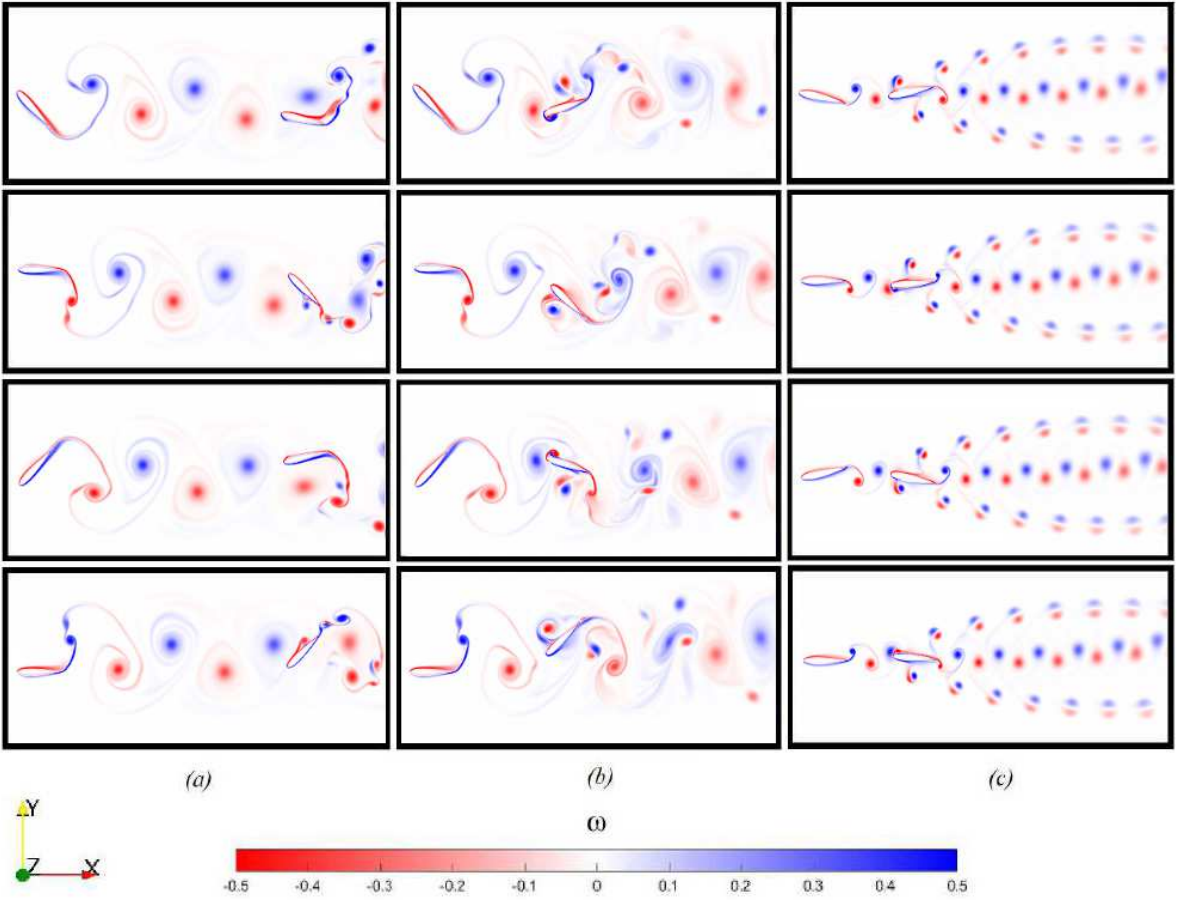}}
  \caption{Snapshots of normalised vorticity for lift cancelling tandem configurations at cycle increments of $t/T =1/4$ where $T=1/f$. Three distinct wake patterns can be observed: (a) Type I at $St_C=0.625$, $\mathcal{S}_C =5$ and $\varphi=1.6 \pi$, (b) Type II at $St_C=0.625$, $\mathcal{S}_C =2$ and $\varphi=1.375 \pi$ and (c) Type III at $St_C=1.5625$, $\mathcal{S}_C=1$ and $\varphi=0.875 \pi$.}
\label{fig:5}
\end{figure}

The manner in which the lift is cancelled depends on the interaction between the back foil and the incoming wake of the front foil. The specific type of interaction affects both the front and the back wake and has a crucial effect on the system‘s overall thrust (see table \ref{tab:2}).  In figure \ref{fig:5}, three possible modes of lift cancellation are reported:

\begin{itemize}
\item Type I, where the back foil slides within the channel between the two vortices that form the incoming dipole (see figure \ref{fig:5}a). 

\item Type II, where the back foil collides with one of the two vortical components of the dipole (see figure \ref{fig:5}b). 

\item Type III, which is effectively an intermediate condition between the previous two modes as the aft foil partly collides with the outer region of the incoming vortex (see figure \ref{fig:5}c). 
\end{itemize}

Interestingly, figure \ref{fig:5}a shows that at $St_C=0.625$ lift cancellation is still present even for a $\mathcal{S}_C \sim 5$. In contrast, most published work supports the idea that downstream flow has no impact on the propulsive characteristics of the fore foil. These studies, however, focus on the interaction between non deflected jets where momentum exchange among subsequent vortices and the ambient fluid is minimal. 

\begin{table}
\begin{center}
\def~{\hphantom{0}}
 \begin{tabular}{c c c c c c c c c} 
 Test Case  & $St_C$ & $A_C$  & $\mathcal{S}_C$  & $\varphi$  & $\overline{C}_{t,f}$  & $\overline{C}_{t,b}$  &$\overline{C}_{l,s
 f}$  &$\overline{C}_{l,b}$ \\ [3pt] 
 Single Foil (Symmetric)& 0.2000 &2.467& - & -& 0.535& - & 0.000 & -\\
  Single Foil (Deflected) & 0.6250 &1.405& - & -& 0.590& - & 0.400 & -\\ 
 Tandem  Foils (Classic)& 0.2000 &2.467& 5.0 & $0.500\pi$& 0.535& 0.910 & 0.002 & 0.014 \\
 Tandem Foils (Type I) & 0.6250 &1.405& 5.0 & $1.600\pi$ & 0.969 & 3.076 & -0.019 & 0.022\\
  Tandem  Foils (Classic) & 0.2000 &2.467& 2.0 & $1.500\pi$& 0.535& 0.909 & -0.017 & 0.033 \\
  Tandem Foils (Type II) & 0.6250 &1.405& 2.0 & $1.375\pi$ & 0.646& -1.191 & 0.039 & -0.013\\ 
   Tandem  Foils (Classic) & 0.2000 &2.467& 1.0 & $0.000\pi$& 0.535& 0.828 & -0.003 & 0.010 \\
  Tandem Foils (Type III) & 1.5625 &0.326& 1.0 & $0.875\pi$ & 0.595& 0.340 & 0.030 & 0.002\\
  \end{tabular}
 \caption{Mean force coefficients $\overline{C}_{t}$ and $\overline{C}_{l}$ of single and tandem flapping foils for kinematics resulting in  symmetric and deflected wakes. Tandem combinations of non deflected wakes are referred to as $\emph{classic}$. }
\label{tab:2}
\end{center}
\end{table}

\subsection{Thrust considerations}
Thrust augmentation is a well reported phenomenon of in line flappers. When the back foil weaves between the incoming vortices, it experiences a higher $U_{\infty}$ compared to the front foil. This increases its thrust generation capacity and cases of $\overline{C}_{t,b} \sim 2 \overline{C}_{t,f}$ have been observed \citep{Muscutt2017}. Table \ref{tab:2} shows that Type I wake modes manage to exceed these values reaching up to $\overline{C}_{t,b} \sim 2.7 \overline{C}_{t,f}$. This should be expected since Type I occurs at a much higher $St_C$ compared to symmetric wake cases found in literature. Hence the circulation of the wake vortices experienced by the back foil is greater enabling the formation of similarly increased strength vortices by the foil and thereby a higher thrust augmentation.

\begin{figure}
  \centerline{\includegraphics[width =4.5 in]{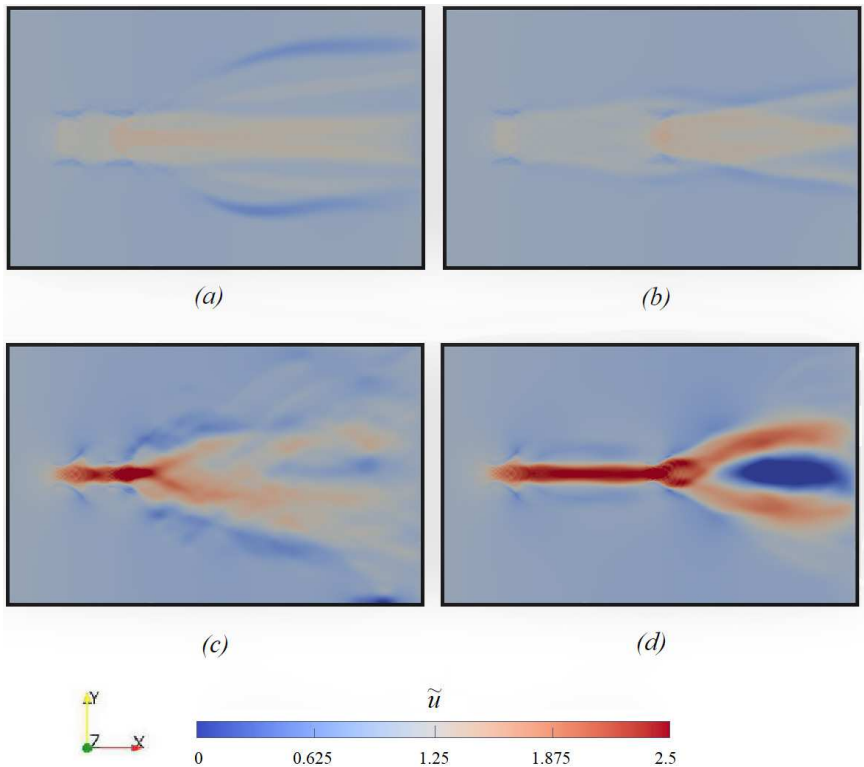}}
  \caption{Time-averaged streamwise velocity  $\tilde{u} = \overline{u}_x / U_{\infty}$ of tandem configurations at $\mathcal{S}_C=1$ (left column) and $\mathcal{S}_C=5$ (right column). Patterns at (a) and (b) show results for high thrust enhancement for a non-deflected wake case with $St_C=0.2$ while (c) and (d) represent type I wake modes with $St_C=0.625$.}
\label{fig:6}
\end{figure}

Figure \ref{fig:6} shows the differences in the time averaged streamwise velocity between high performance conventional cases at $St_C=0,2$ and Type I modes at $St_C=0.625$. Conventional test cases utilize the optimal combinations $\varphi=0~\pi,~\mathcal{S}_C = 1$ and $\varphi=0.5~\pi,~\mathcal{S}_C = 5$ derived from the work of \cite{Muscutt2017} while Type I wake modes are derived for $\varphi=1~\pi,~\mathcal{S}_C = 1$ and $\varphi=1.6~\pi,~\mathcal{S}_C = 5$ respectively. Clearly, the peak value of $\overline{u}_{x}/U_{\infty}$  is much higher within Type I wakes due to the higher circulation of the vortices. Furthermore, the comparatively shorter distance between these vortices leads to a much narrower jet. Since $T=\int u^2 dy$ , this condition enables the higher values of thrust reported above.

\subsection{The physics of lift cancellation}
\begin{figure}
  \centerline{\includegraphics[width =5.5 in]{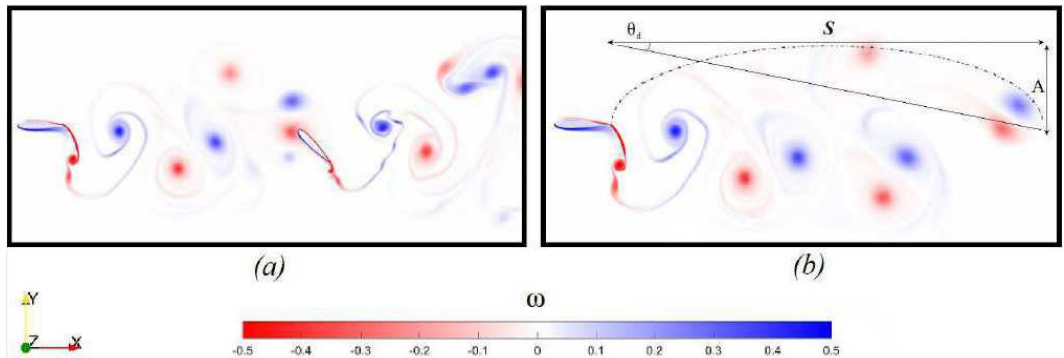}}
  \caption{Normalised vorticity field of tandem (a) and single (b) foils at $A_C=1.405 $ and $St_C=0.625$. The initial dipole of a single foil travels in an elliptical path (dashed line) which enhances deflection during its final stages. However, the presence of a downstream flapping foil at an angle $\geq \theta_d$ leads to the decomposition of this first dipole, forcing the incoming jet to maintain symmetry.}
\label{fig:7}
\end{figure}

The mechanism of lift cancellation is depicted in figure  \ref{fig:7}a. The initial dipole shed from the leading foil advects downstream and splits in to two after colliding with the vertically moving aft foil. Consequently, the advection speed of the clockwise vortex (red) decreases which reduces its distance from the subsequent dipole shed by the front foil. This affects the cohesion of the second dipole and its clockwise vortex (blue) is now under the influence of two counter-clockwise vortices. This situation propagates upstream, affecting every subsequent dipole in the same manner, resulting in wake convergence towards the centreline forming a classic reverse $\emph{von}~K\acute{a}rm\acute{a}n$  street. 

As the collision of the first dipole with the rear foil is necessary for the lift cancellation, we need to examine the circumstances that would lead to this interception. By definition, a collision between two entities is only possible if their paths intersect. Assuming that the first dipole travels along an inclined downstream path set by the initial advection speed (see figure \ref{fig:7}b), it can be shown that this path diverts from the freestream at an angle of $tan \theta_d$  $\propto$ $fA/U_\infty$ (or  $\tan \theta_d = p fA/U_\infty$). This is based on assuming that the horizontal advection speed is proportional to the freestream speed ($U_\infty$) and the vertical advection speed is proportional to the vertical speed of the trailing edge of the front foil ($fA$). Note that the actual path of the dipole is not a straight line as it follows a more complex elliptic path. However, the factors leading to the overall angle are sufficient for this discussion. To achieve lift cancellation, this angle, $\tan \theta_d$, must be smaller than the largest angle between the trailing edge of the front foil and the L.E. of the back foil, $\tan \theta_g = A/\mathcal{S}$ so that:
\begin{equation}
\begin{array}{l}
\displaystyle ~~~\tan \theta_d \leq \tan \theta_g \,\to\, p f A / U_\infty \leq A/\mathcal{S} \\
\\
\displaystyle  ~~~~~~~~~~~~~ \therefore f \mathcal{S} /U_\infty  \leq 1/p\\
\end{array}
\label{eqo}
\end{equation}
This suggests that there is a $spacing$ $based$ Strouhal number that will act as a clear boundary between areas where lift cancellation is possible and areas where wake deflection is maintained. This geometric relationship accounts for all possible phase differences between the fore and aft foils and could even be independent of the frequency of the aft foil. This Strouhal number only depends on the ratio between the horizontal and vertical advection speeds of the dipole shed by the front foil which may vary with the kinematics of the front foil.

\begin{figure}
  \centerline{\includegraphics[width = 2.8 in]{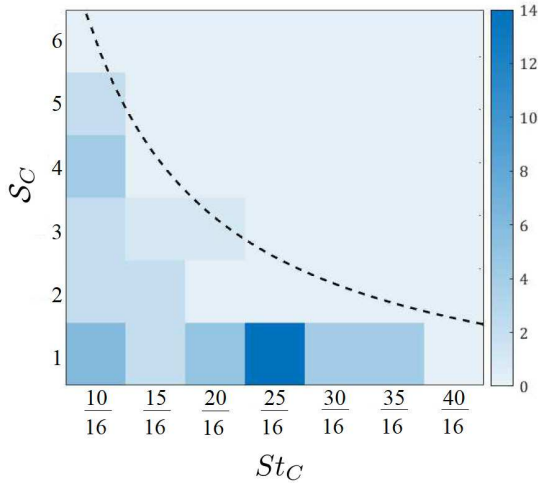}}
  \caption{Heatmap of lift cancellation instances for a tandem foil system and the full range of $\varphi \sim [0\pi - 1.75\pi]$ and various $St_C - \mathcal{S}_C$ combinations. The dashed curve marks the boundary of possible lift cancellation occurrence and corresponds to a $spacing$ $based$ Strouhal $St_{\mathcal{S}} = f  \mathcal{S} / U_{\infty} = 4$. }
\label{fig:8}
\end{figure}

The number of lift cancellation instances are plotted on the map of figure \ref{fig:8} as a function of non-dimensional spacing ($\mathcal{S}_C$) and non-dimensional frequency ($f\mathcal{C}/U$). We observe that lift cancellation is impossible above a certain region marked with dashed solid black curve. This curve has the form $St_C * \mathcal{S}_C = f \mathcal{S} /U_\infty = const.$. Fitting this equation to the data in figure \ref{fig:8} gives a $spacing-based$ Strouhal number that determines the lift cancellation border at $St_\mathcal{S} =f \mathcal{S} / U_{\infty} \sim 4$. Taking into account \ref{eqo} this means that $p=1/4$. In other words, for a given spacing $\mathcal{S_C}=1$ the back foil has the opportunity to impose wake symmetry only if its vertical speed is approximately less than (or equal to) a quarter of the horizontal advection speed of the first shed dipole. 

\begin{figure}
  \centerline{\includegraphics[width =5.5 in]{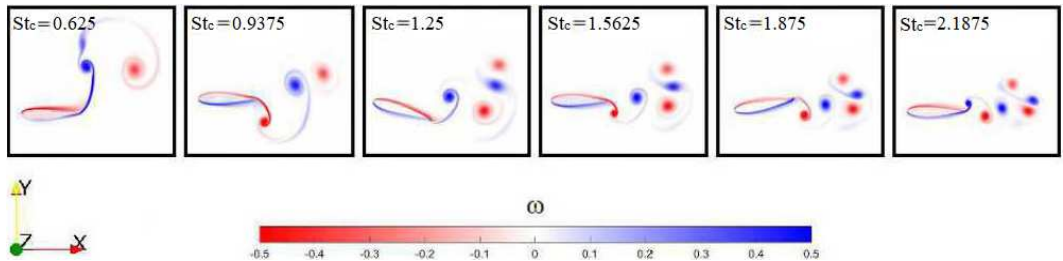}}
  \caption{Early stages of wake development at $\mathcal{S}_C=1$ downstream of a single flapping foil.}
\label{fig:9}
\end{figure}

A careful look at figure \ref{fig:8} reveals that the highest concentration of lift cancellations occurs at $\mathcal{S}_C = 1$ and $St_C= 25/16 =1.5625$. Once again this is linked to the development and propagation of the first permanent dipole. Figure \ref{fig:9} shows that at $St_C = 1.5625$ and $\mathcal{S}_C = 1$ the path of the dipole coincides with the centreline. In addition, the channel between its two vortices is approximately equal to the thickness of the back foil. This is an ideal condition, as it maximises the chances of collision with the dipole and thus the number of $\varphi$ that lead to lift cancellation. Any increase in the $St_C$ changes the direction of the dipole and narrows the channel between its vortices, making it harder to dissolve by the back foil. On the other hand, a lower frequency leads to underdeveloped and more disoriented dipoles, minimizing the chances of impact. Consequently the amount of $\varphi$ able to eliminate $\overline{C}_l$ is reduced and with it, the amount of lift cancellation instances. This essentially describes a Goldilocks condition where the $St_C = 1.5625 , \mathcal{S}_C = 1$ is the optimal combination for lift cancellation throughout the tested parameter space.
i

\section{Conclusions}
Flapping foils, generating fully deflected wakes, are analysed in both single and tandem configurations undergoing coupled heave and pitch kinematics. We find that deflected wakes generated by foils undergoing coupled heave and pitch motion are similar to those reported for pure heave or pure pitch cases. Likewise, the driving mechanism of deflection is the development of dipole structures and their subsequent departure from the centreline. 

Tandem configurations are able to reorder deflected jets into symmetric wakes with $\overline{C}_{l,f} = \overline{C}_{l,b} \sim 0$. Certain $\varphi~,~\mathcal{S}_C~,~St_C$ combinations are shown to direct both wakes back to the centreline even for inter foil distances of 5 chord lengths. To achieve lift cancellation the back foil has to dissolve the first shed dipole of the front wake and this is achieved when the angle between the T.E. of the front foil and the L.E. of the back is greater than the dipole's convection angle. This can be expressed via a maximum $spacing~based~Srouhal$, $St_{\mathcal{S}} = 4$, above which any lift cancellation is impossible. Physically, this $St_{\mathcal{S}}$ limit implies that the back foil must encounter the upstream wake within a few motion cycles for lift cancellation to be possible. 

When the total lift of the tandem configuration is cancelled, the wake downstream becomes symmetric. Three wake modes are reported. Type I occurs when the back foil separates the two vortical components of the incoming dipole by weaving between them which leads to a remarkable increase in $\overline C_{t}$. Type II mode occurs when the L.E. directly collides on one of the dipole's vortices which introduces a significant drag penalty. In addition, an intermediate mode Type III is reported whose behaviour varies according to the intensity of the vortex-foil collision.

This study is the first to provide evidence of the significant impact of the downstream field to the front foil. Furthermore, it is demonstrated that the wake deflection can be diminished with a subsequent remarkable thrust enhancement. These findings can support the design of high performance biomimetic propulsors, as a simple change of the back foil's phase enables high thrust generation without side force at high frequencies previously considered impossible. 

\section*{Acknowledgements}
This research was supported financially by the Office of Naval Research award
no. N62909-18-1-2091 and the Engineering and Physical Sciences Research Council
doctoral training award (1789955). All data and post processing scripts supporting this study are openly available via the University of Southampton repository at https://doi.org/10.5258/SOTON/D1397. 

\bibliographystyle{jfm}
\bibliography{jfm-instructions}

\end{document}